\documentclass[prl,onecolumn, superscriptaddress,amsmath,amssymb, aps,10pt
]{revtex4-2}

\usepackage{graphicx}% Include figure files
\usepackage{dcolumn}% Align table columns on decimal point
\usepackage{bm}% bold math
\usepackage{braket} %braket
\usepackage{xcolor}
\usepackage{xr}
\makeatletter
\newcommand*{\addFileDependency}[1]{
  \typeout{(#1)}
  \@addtofilelist{#1}
  \IfFileExists{#1}{}{\typeout{No file #1.}}
}
\makeatother

\newcommand*{\myexternaldocument}[1]{
    \externaldocument{#1}
    \addFileDependency{#1.tex}
    \addFileDependency{#1.aux}
}
\myexternaldocument{SM}%
\usepackage{hyperref}% add hypertext capabilities
\DeclareUnicodeCharacter{2212}{-}

\renewcommand{\baselinestretch}{1.5}

\begin{document}
%\preprint{APS/123-QED}

\title{How to use the dispersion in the $\chi^{(3)}$ tensor for broadband generation of polarization-entangled photons}

\author{Valeria Vento}
\affiliation{Institute of Physics, Swiss Federal Institute of Technology Lausanne (EPFL), CH-1015 Lausanne, Switzerland} 
\affiliation{Center of Quantum Science and Engineering, Swiss Federal Institute of Technology Lausanne (EPFL), CH-1015 Lausanne, Switzerland
}%
\author{Francesco Ciccarello}
\affiliation{Institute of Physics, Swiss Federal Institute of Technology Lausanne (EPFL), CH-1015 Lausanne, Switzerland} 
\affiliation{Center of Quantum Science and Engineering, Swiss Federal Institute of Technology Lausanne (EPFL), CH-1015 Lausanne, Switzerland
}%
\author{Sakthi Pryia Amirtharaj}
\affiliation{Institute of Physics, Swiss Federal Institute of Technology Lausanne (EPFL), CH-1015 Lausanne, Switzerland} 
\affiliation{Centre for Quantum Information, Communication and Computing, Indian Institute of Technology Madras, Chennai, Tamil Nadu 600036, India
}%
\author{Christophe Galland}
\email{chris.galland@epfl.ch}
\affiliation{Institute of Physics, Swiss Federal Institute of Technology Lausanne (EPFL), CH-1015 Lausanne, Switzerland} 
\affiliation{Center of Quantum Science and Engineering, Swiss Federal Institute of Technology Lausanne (EPFL), CH-1015 Lausanne, Switzerland
}%
\date{\today}% It is always \today, today,
\begin{abstract}

%\textbf{[Francesco abstract for conference - to be modified]Two-photon coincidence measurements under pulsed excitation offer various ways to study the non-linear response of materials, including Raman scattering. Emission of correlated Stokes – anti-Stokes photon pairs can be mediated by the exchange of real \cite{} or virtual \cite{} phonons, a process described by a resonant term in the third-order $\chi^{(3)}$ tensor of the material. When the non-resonant, electronic contribution to $\chi^{(3)}$ is added, the quantum state of photon pairs emitted by spontaneous four wave mixing can become polarization entangled at some Raman shifts \cite{}. Here, we perform a spectrally-resolved Bell test on single-crystal diamond pumped by 1 ps laser pulses, using fiber dispersion and two-photon coincidence measurements. Our results show a statistically significant violation of the CHSH Bell inequality in a large range of Raman shifts around the pump, in agreement with the theoretical predictions.}
\textbf{
Polarization-entangled photon pairs are a widely used resource in quantum optics and technologies, and are often produced using a nonlinear process. Most sources based on spontaneous parametric downconversion have relatively narrow optical bandwidth because the pump, signal and idler frequencies must satisfy a phase-matching condition. Extending the bandwidth, for example to achieve spectral multiplexing, requires changing some experimental parameters such as temperature, crystal angle, poling period, etc. Here, we demonstrate broadband (tens of THz for each photon) generation of polarization-entangled photon pairs by spontaneous four-wave mixing in a diamond crystal, with a simple colinear geometry requiring no further optical engineering. Our approach leverages the quantum interference between electronic and vibrational contributions to the $\chi^{(3)}$ tensor. Entanglement is characterized in a single realization of a Bell test over the entire bandwidth using fiber dispersion spectroscopy and fast single-photon detectors. The results agree with the biphoton wavefunction predicted from the knowledge of the $\chi^{(3)}$ and Raman tensors and demonstrate the general applicability of our approach to other crystalline materials.} 

\end{abstract}

\maketitle

\pagebreak
\textit{Introduction.---} 
Nonlinear optical processes play a pivotal role in producing and engineering quantum states of light. Spontaneous parametric-down conversion (PDC) and four-wave mixing (FWM), which rely respectively on the $\chi^{(2)}$ and $\chi^{(3)}$ susceptibility of materials, have been employed for generating a variety of photonic quantum states \cite{wang_integrated_2021} -- entangled photon pairs generated in the low-gain regime being the most popular class.
%These two kind of processes are used in different contexts ....
%tunability of the properties of the 2-photon source: directionality, spectrum and polarizaiton.
PDC sources generally offer the highest brightness because, for non-centrosymmetric crystals, $\chi^{(2)}$ is much larger than $\chi^{(3)}$. A challenge with PDC is the large frequency difference between the pump and the signal/idler fields that restricts the spectral bandwidth over which phase matching can be achieved without modifying the experimental conditions or implementing delicate dispersion engineering in periodically poled waveguides \cite{javid2021}. Moreover, materials such as silicon and diamond have no $\chi^{(2)}$ nonlinearity while being popular platforms for integrated quantum photonics at telecom and visible wavelengths, respectively \cite{moody2022}; it motivates the use of FWM for generating entangled photons.

Among the photonic degrees of freedom that can be entangled, polarization is often the easiest to manipulate and measure. %Accordingly, polarization-entangled photon pairs have become a fundamental resource for quantum computation and communication technologies \cite{XXX}. 
However, generating polarization entanglement usually requires careful engineering of the source, since parametric nonlinear processes more naturally generate photon pairs that are entangled in the temporal and spatial degrees of freedom due to energy and momentum conservation, respectively. 
There are relatively few demonstrations of polarization-entangled photon pair produced by spontaneous FWM, with the most established methods relying on the $\chi^{(3)}$ nonlinearity of optical fibers \cite{takesue2004,li2005,fan2007,fang_multidimensional_2016} or silicon waveguides \cite{takesue2008} that are themselves embedded within a Sagnac loop. The photon pairs produced by FWM in glass fibers or silicon waveguides are typically copolarized, so that entanglement is generated from two orthogonal sources after erasing distinguishing information in all other degrees of freedom (spatial and temporal). 
This erasure has also been achieved using a fully-integrated silicon photonic circuit comprising a polarization rotator between TE and TM modes \cite{matsuda2012}. 
In an effort to simplify such FWM sources, researchers leveraged the tensorial nature of the $\chi^{(3)}$ nonlinear response and engineered the polarization mode dispersion to achieve polarization entanglement directly out of silicon \cite{Lv2013} and AlGaAs \cite{kultavewuti2017} waveguides. 

When FWM is used to generate photon pairs, vibrational Raman scattering is mostly known as a source of noise degrading the purity of the joint quantum state \cite{inoue2004}. This problem is particularly pronounced in amorphous media such as optical fibers because the Raman scattered light has a broad spectrum \cite{agrawal2007}. In crystalline materials such as silicon, AlGaAs and diamond, the Raman peak is very sharp (few wavenumbers), so that Raman noise is more easily avoided by choosing an appropriate detuning from the pump. Interestingly, a recent work by Freitas \textit{et al.} showed that polarization entanglement can be generated from bulk diamond for pairs detuned by $\pm 900\text{~cm}^{-1}$ from the degenerate two-photon pump \cite{freitas_microscopic_2023}, a result qualitatively attributed to the interference of electronic and vibrational (Raman) microscopic pathways leading to photon pair creation. 

In this Letter, we establish a theory of polarization-entangled photon pair generation using FWM in diamond (also applicable to silicon and other crystals for which the $\chi^{(3)}$ and Raman tensors are known) and validate our predictions by performing coincidence measurements and Bell tests over a large optical bandwidth of $2200\text{~cm}^{-1}$ ($65$~THz) on either side of the pump frequency (spanning 130 THz in total). Moreover, we achieve this characterization in a single set of polarization measurements by employing fiber spectroscopy \cite{avenhaus_fiber-assisted_2009}, where the frequency information of individual photons is mapped onto detection time through the natural dispersion of optical fibers. The relaxed phase-matching condition for FWM enables statistically significant violation of the CHSH inequality \cite{clauser_proposed_1969} for the $\Psi^+$ Bell state across a considerable bandwidth, at all measured frequency shifts below $1250\text{~cm}^{-1}$ ($37.4$~THz) -- a bandwidth on par with that of state-of-the-art single-photon sources in optimized photonic platforms \cite{lim2008,alshowkan2022}.
%While recent studies have shown frequency-dependent entanglement originating from diamond \cite{freitas_microscopic_2023}, a theoretical framework elucidating the data based on the material's nonlinearity remained absent. 
We find that our results agree with the theoretical predictions accounting for the frequency dependence of $\chi^{(3)}$ that is due to the sharp vibrational Raman resonance at $1332\text{~cm}^{-1}$. 

We notice that the term “broadband" is sometimes used in the literature to indicate correlated photon-pair sources capable of establishing entanglement between far-detuned photons -- typically to bridge visible and telecom ranges -- while keeping a narrow linewidth \cite{duan_broadband_2024}. On the contrary, we generate a broad bandwidth of entangled photon pairs at once with a single pump laser of fixed frequency. In this sense, our technique also offers an advantage over coherent anti-Stokes Raman scattering (CARS) for probing third-order nonlinearities. Indeed, the dispersion features we analyze could be obtained with CARS only by employing an additional broadly tunable ($170$~nm) visible pump. In analogy to CARS, a recent work \cite{zheng_phonon-mediated_2024} used a tunable THz source to evidence third-harmonic enhancement and suppression in diamond, qualitatively explained in terms of the same $\chi^{(3)}$ dispersion.

\begin{figure}[t]
\centering
\includegraphics[width=0.45\textwidth]{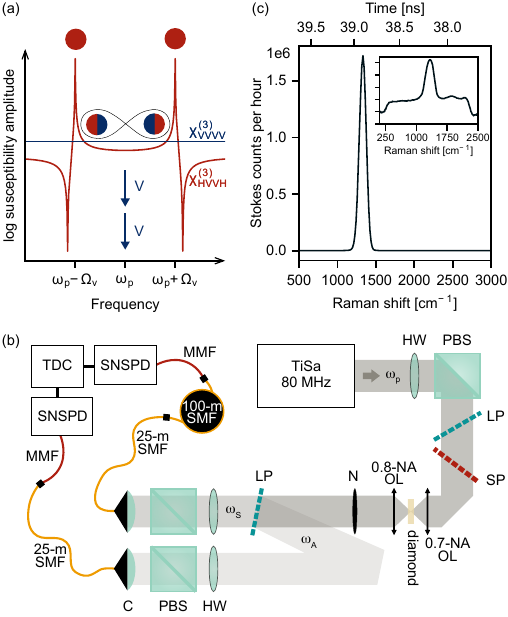}
\renewcommand{\baselinestretch}{1.1}
\caption{ (a) Sketched amplitude of the two relevant $\chi^{(3)}$ tensor elements in diamond under vertically polarized pump (blue arrows) at $\omega_p$; $\Omega_\nu$ is the Raman-active optical phonon frequency. Circles of matching colors represent correlated photons at Stokes ($\omega_S$) and anti-Stokes ($\omega_A$) frequencies, with the generation of the $\Psi^\pm$ Bell states expected where $|\chi^{(3)}_{HVVH}| \simeq |\chi^{(3)}_{VVVV}|$. %The polarization state of the correlated pair is determined by $\chi^{(3)}_{HVVH}$ (red curve) and $\chi^{(3)}_{VVVV}$ (blue curve). 
(b) Experimental setup. HW: half-wave plate, PBS; polarizing beam splitter, OL: objective lens, LP: tunable long-pass filter, SP: tunable short-pass filter, N: notch, C: coupler, SMF: single-mode fiber, MMF: multimode fiber, SNSPD: superconducting nanowire single-photon detector (Eos 210 CS, Single Quantum), TDC: time-to-digital converter (ID900, ID Quantique). The sample is a $300$-$\mu$m thick synthetic diamond crystal cut perpendicular to the [100] crystal axis.  (c) Stokes spectrum measured by single-photon fiber spectroscopy (logarithmic scale in inset).  }
\label{fig:setup}
\end{figure}
%Tunable long-pass (LP) and short-pass (SP) filters are used to cut the pump spectral width (increase the pulse duration).
%Stokes spectrum obtained by converting the arrival time of the Stokes photons (stop signal) after the laser synchronization pulse (start signal) into wavenumbers. The inset shows the same spectrum in logarithmic scale

\vspace{10mm}
\textit{Results.---} 
Figure~\ref{fig:setup}a illustrates the concept of the experiment in the frequency domain. Two vertically-polarized pump photons at frequency $\omega_p$ can generate pairs of correlated photon pairs at $\omega_S$ (Stokes) and $\omega_A\equiv 2\omega_p-\omega_S$ (anti-Stokes) frequencies, satisfying energy conservation. The polarization state of the correlated pair is determined by the third-order susceptibility tensor of diamond, as we will explicate below. Quite generally, the existence of a Raman-active optical phonon mode at frequency $\Omega_\nu$ causes a resonant contribution to the $\chi^{(3)}$ tensor when photon pairs satisfy $\omega_{A/S}=\omega_p \pm \Omega_\nu$ \cite{boydNLO}. The aim of this article is to show how these intrinsic material properties can be used to produce polarization-entangled photons over a large bandwidth around $\omega_p$ without any further optical engineering. Our results are therefore applicable to other crystalline materials. 

Figure~\ref{fig:setup}b shows the experimental setup.
A mode-locked Titanium:Sapphire oscillator (Tsunami, Spectra Physics) generates $200$-fs pulses at $781$~nm with a repetition rate of $80$~MHz. In order to resolve the sharp spectral dependence of the susceptibility sketched in Fig.~\ref{fig:setup}a, the pump pulse duration should be of the same order of magnitude as the phonon coherence time, which is around 8~ps in diamond \cite{tarragovelez2020}. We use tunable long-pass (LP) and short-pass (SP) filters to increase the pulse duration to $1.1$~ps by reducing the spectral width to $0.8$~nm. %The strongly focused excitation/detection geometry relaxes the phase matching conditions required for nonlinear processes to occur. 
The average power on the sample is $48$~mW (600~pJ pulse energy), below the onset of single-beam stimulated processes \cite{vento_measurement-induced_2023}. 

The collected Stokes and anti-Stokes signals are split into two different paths by a longpass interference filter placed almost perpendicular to the beam to minimize the polarization-dependent response between $s$ and $p$ components. %, particularly pronounced on the transmitted beam when using a conventional 45$^\circ$ geometry. 
Next, each signal is spectrally dispersed through $25$~m of single-mode fibers (SMFs). Since losses also increase with the fiber length, we add $100$~m SMF only to the Stokes path. Knowing the dispersion parameters of the fibers, we can convert photon detection times into detuning from the pump in wavenumbers \cite{supp}. Superconducting nanowire single-photon detectors (SNSPDs) allow us to achieve a temporal resolution of around $30$~ps per detector (time jitter). Figure~\ref{fig:setup}c shows the Stokes spectrum obtained the laser synchronization pulse is used as the start signal and the detection of a $H$ polarized Stokes photon as the stop signal. 

Fiber spectroscopy had been employed to measure various spectral properties of photon pairs generated by PDC, like the joint spectral amplitude \cite{zielnicki_joint_2018} and the second-order correlation function \cite{okoth_microscale_2019}. Ref.~\cite{weissflog_tunable_2023} measured the spectrum of PDC-generated polarization-entangled photons, but the spectral variation of the quantum state was not addressed. Our work reveals the frequency evolution of the degree of purity, polarization entanglement, and non-locality of the biphoton state generated by FWM. Our measurement technique can also be used to characterize novel sources of polarization-entangled photons implemented in flat optical systems \cite{sultanov_flat-optics_2022, sharapova_nonlinear_2023, santos_entangled_2024}, including metamaterials \cite{santiago-cruz_resonant_2022,ma_polarization_2023} and nanoresonators \cite{lee_photonpair_2017, weissflog_nonlinear_2024}.

\begin{figure}[h!]
\centering
\includegraphics[width=0.4\textwidth]{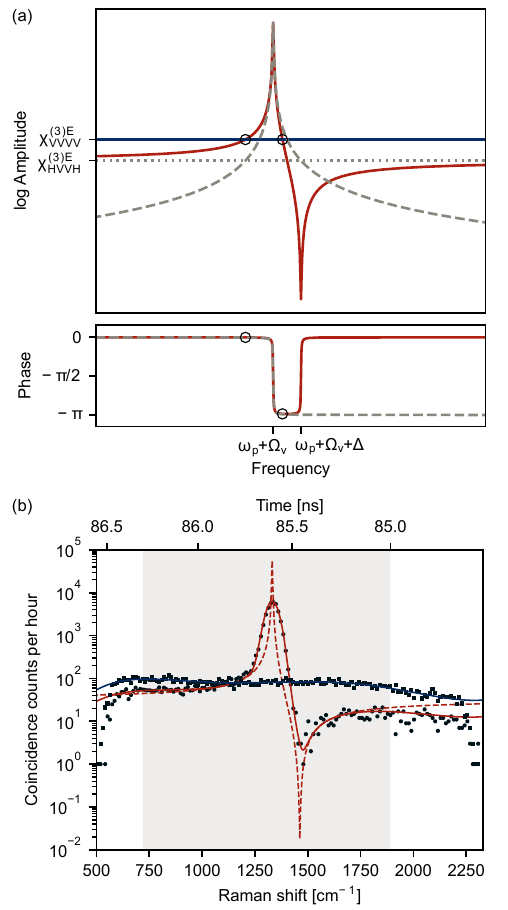}
\renewcommand{\baselinestretch}{1.1}
\caption{ (a) Theoretical amplitude and phase of $\chi^{(3)}_{VVVV}$ (solid blue) and $\chi^{(3)}_{HVVH}$ (solid red). The Raman and the electronic contributions to the latter are shown as dashed and dotted gray lines respectively. At the frequencies where $\chi^{(3)}_{VVVV}$ and $\chi^{(3)}_{HVVH}$ have the same amplitude (black circles), the maximally entangled states $\ket{\Psi_\pm}$ are generated (Eq.~\ref{eq:Psipm}). (b) Frequency-resolved anti-Stokes--Stokes coincidences measured after selecting $V$ (black squares) or $H$ (black circles) polarization in the two detection paths. The data are acquired in $1$ hour. The solid blue curve is a polynomial fit that reflects the setup polarization response function since no resonances are expected in this configuration. The solid red curve represents the fit function $A\log(I_H(\omega)\ast G(\omega))+b$ multiplied by the setup response, where the free parameters $\Delta = 130.59 \text{~cm}^{-1}$, $\sigma_G=24.80 \text{~cm}^{-1}$, $A=0.82$, $b=1.51$ are extracted by fitting within the gray region. The dashed red line shows the fit function without convolution, corresponding to the red line in panel (a).}
\label{fig:coinc}
\end{figure}

Since diamond is centrosymmetric, the third-order susceptibility mediates the lowest-order nonlinear optical processes. It is a complex rank-3 tensor that depends of the frequencies and polarizations of the four fields involved: $\chi^{(3)}_{ijkl}(\omega_A,\omega_p,\omega_p,\omega_S)$, where the indices $i,j,k,l$ specify the polarization directions of the fields at the four frequencies listed in the same order in the parenthesis. In a stimulated scattering experiment, two incoming beams at $\omega_p$ and $\omega_S$ can generate light at $\omega_A$ through the following third-order nonlinear processes: (i) stimulated Raman scattering, when $\omega_p-\omega_S=\Omega_\nu$, which is the optical phonon frequency (ii) two-photon absorption, when $2\omega_p$ corresponds to an electronic transition, and (iii) any non-resonant FWM process mediated by far-detuned electronic levels. In 1974, Levenson and Bloembergen performed stimulated FWM spectroscopy on diamond and from the independent knowledge of the Raman tensor and Raman gain they extracted all components of its $\chi^{(3)}$ tensor \cite{levenson_dispersion_1974}. 

%For such processes to occur, phase-matching conditions must be satisfied, which result in the generation of correlated photon pairs at $\omega_S$ and $\omega_A$. 
In our experiment, we employ a single pump beam with frequency $\omega_p$ far from any electronic resonance. We can therefore neglect two-photon absorption and write $\chi^{(3)}=\chi^{(3)\text{E}}+\chi^{(3)\text{R}}$ where $\chi^{(3)\text{E}}$ is the non-resonant -- hence real -- electronic FWM contribution and $\chi^{(3)\text{R}}$ is the (complex) vibrational Raman scattering contribution featuring a resonance at $\omega_{A/S}=\omega_p \pm \Omega_\nu$, \cite{boydNLO} where $\Omega_\nu$ is the frequency of the optical phonon at $1332\text{~cm}^{-1}$. The pump propagates along the [100] crystallographic axis and the diamond is rotated to align the [001] axis with the pump polarization, which is vertical in the laboratory frame \cite{supp}. In this configuration, Raman scattering ($\chi^{(3)\text{R}}$) only generates horizontally polarized pairs \cite{levenson_dispersion_1974,freitas_microscopic_2023}. 
Figure~\ref{fig:coinc}a shows the amplitude and phase of the non-zero components of the susceptibility tensor contributing in this case
\begin{align}
    \chi_{VVVV}^{(3)} &= \chi_{VVVV}^{(3)\text{E}}\equiv F\chi_{HVVH}^{(3)\text{E}} \label{chi_VVVV}\\
    \chi_{HVVH}^{(3)} &= \chi_{HVVH}^{(3)\text{E}}+\chi^{(3)\text{R}}=\chi_{HVVH}^{(3)\text{E}}\left(1+\frac{\Delta}{\Omega_\nu-\omega + i \Gamma_\nu} \right) \label{chi_HVVH}
\end{align}
where $\omega=\omega_p-\omega_S=\omega_A-\omega_p$, and $\Gamma_\nu=2\pi c \gamma_\nu$ is the vibrational decay rate with $\gamma_\nu\simeq 1  \text{~cm}^{-1}$ the Raman linewidth. Following Ref.\cite{levenson_dispersion_1974}, we define $\Delta=\frac{|\alpha^R|^2N}{24\hbar\chi_{HVVH}^{(3)\text{E}}}$, where $\alpha^R$ is proportional to the Raman tensor and $N$ is the volumic density of unit cells. The value of $\Delta$ corresponds to the angular frequency difference between the maximum and minimum of $|\chi_{HVVH}^{(3)}|$, i.e. between the points of maximum constructive and destructive interference of $\chi^{(3)\text{E}}$ and $\chi^{(3)\text{R}}$ (Fig.~\ref{fig:coinc}). 

The measured rate of photon pairs for $V$ and $H$ polarizations follow the proportionality relations $I_V(\omega)\propto |\chi_{VVVV}^{(3)}|^2$ and $I_H(\omega) \propto |\chi_{HVVH}^{(3)}|^2$, respectively. We measure them by counting the coincidences between anti-Stokes and Stokes photons after they pass through parallel polarizers, as displayed in Fig.~\ref{fig:coinc}b. Since $\chi_{VVVV}^{(3)}$ is theoretically constant in frequency, we use the measured $I_V(\omega)$ as a characterization of the setup spectral response, which we assume polarization independent (the polarization of photons is largely scrambled by the time they impinge on the SNSPD). To account for the spectral broadening due to the finite pump bandwidth and the temporal jitter of the detectors, the theoretical $H$-polarized intensity $|\chi_{HVVH}^{(3)}|^2$ is convoluted with a Gaussian $G(\omega)$ of variance $\sigma_G^2$ centered at $\Omega_\nu$.  %From the fit in a restricted frequency region around the resonance, we extract $\Delta = 130.59 \text{~cm}^{-1}$ and $\sigma_G=24.80 \text{~cm}^{-1}$. 
After multiplying by the setup response extracted above, the resulting model (red line in Fig.~\ref{fig:coinc}b) accurately reproduces our data. 
%Far from the phonon resonance, we can also extract the parameter $F=\frac{I_V(\omega)}{I_H(\omega)}\big|_\text{off res} \simeq 2$. 
% sentence on the fact that this curve is the same that you would obtain by measuring the anti-Stokes intensity in stimulated scattering experiment

\begin{figure}[h!]
\centering
\includegraphics[width=0.4\textwidth]{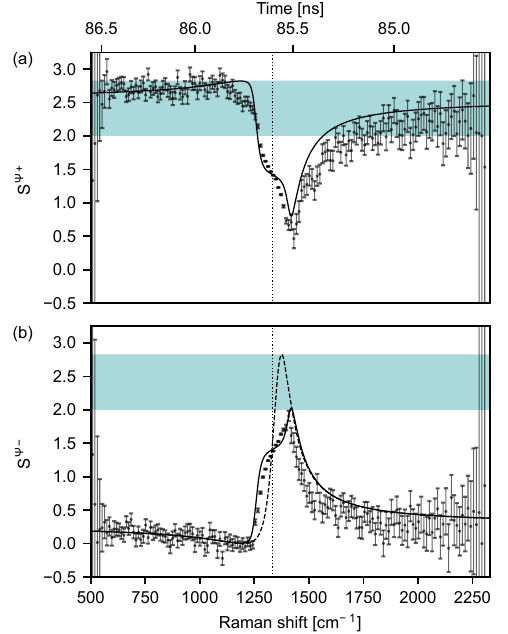}
\renewcommand{\baselinestretch}{1.1}
\caption{(a-b) Symbols represent the measured CHSH parameters $S^{\Psi+}(\omega)$ (a) and $S^{\Psi-}(\omega)$ (b) extracted from a set of $16$ coincidence measurements of $1$ hour each ($4$ measurements for each of the $4$ correlation parameters $E_{\theta_1,\theta_2}(\omega)$ appearing in Eq.~\ref{eq:S}, see also \cite{supp}). The error bars are calculated by propagating the standard deviation of the coincidence counts in the region $[630, 890]\text{~cm}^{-1}$. The shaded domain is delimited by the upper classical bound of $2$ and the upper quantum bound of $2\sqrt{2}$. The dotted black vertical line indicates the Raman resonance, solid lines are theoretical curves computed from Eqs.~\ref{eq:n}-\ref{eq:S}. In (b) the dashed curve is computed without accounting for the setup response.}
%(c) Measured $E_{\theta_1,\theta_2}$ at $910\text{~cm}^{-1}$ (dashed red line in (a)) for $\theta_1=0$ (circles) and $\theta_1=\frac{\pi}{4}$ (triangles) and different values of $\theta_2$. Blue points correspond to the measurements used for the Bell tests in panels (a-b). Solid lines are model predictions.} and the dashed red line is a sinusoidal fit.}
\label{fig:interf}
\end{figure}

In the End Matter, we give the general biphoton polarization state generated under arbitrary pump polarization. 
For a vertical pump polarization the wavefunction reduces to
\begin{equation}
	\ket{\psi(\omega)} = \chi_{HVVH}^{(3)\text{E}} \left[F \ket{VV}+\left(1+\frac{\Delta}{\Omega_\nu-\omega + i \Gamma_\nu}\right)\ket{HH}\right]. \label{eq:state}
\end{equation}
As customary for spontaneous PDC and FWM in the low gain regime, this biphoton wavefunction should be understood as the one-photon-pair component of the full two-mode squeezed state. It is an excellent description of the results obtained by coincidence measurements (which are not sensitive to the vacuum component) as long as the probability of photon pair emission is much smaller than one per pulse so that double-pair emission is a negligible process.
The probability $P_{\theta_1,\theta_2}(\omega)$ to measure a coincidence when Stokes and anti-Stokes polarizers are set at angles $\theta_1$ and $\theta_2$ is expressed in the End Matter; from Eq.~\ref{eq:P} we extract the correlation parameter
 $   E_{\theta_1,\theta_2}(\omega) = P_{\theta_1,\theta_2}(\omega) +P_{\theta_1+\pi/2,\theta_2+\pi/2}(\omega)-P_{\theta_1,\theta_2+\pi/2}(\omega)-P_{\theta_1+\pi/2,\theta_2}(\omega)$
and the CHSH parameters \cite{clauser_proposed_1969}
\begin{equation}
    S^{\Psi\pm}(\omega)=E_{0,\theta}(\omega)+E_{0,-\theta}(\omega)\pm E_{2\theta,\theta}(\omega)\mp E_{2\theta,-\theta}(\omega)\label{eq:S}
\end{equation}
where $\theta=\pi/8$. The corresponding Bell inequalities
$S^{\Psi\pm}(\omega) \leq 2$
are maximally violated by the Bell states 
\begin{equation}
    \ket{\Psi_\pm}=\frac{\ket{HH}\pm\ket{VV}}{\sqrt{2}} \label{eq:Psipm}
\end{equation}
respectively.
%Any maximally entangled state maximally violates some Bell inequality. Therefore, the choice of the CHSH parameter depends on the maximally entangled state that we want to measure.In the case of vertical pump polarization, the susceptibility components $\chi^{(3)}_{VVVV}$ and $\chi^{(3)}_{HVVH}$ generate the maximum entangled states at the two frequency points highlighted in Fig.~\ref{fig:coinc}(a).
As shown in Figures \ref{fig:interf}a-b, we extract the parameters $S^{\Psi\pm}(\omega)$ for $\phi=\xi=0$ from the direct measurement of $n_{\theta_1,\theta_2}(\omega)$ and compare them with the values theoretically computed from Equations \ref{eq:n}-\ref{eq:S} with $\Delta$ and $\sigma_G$ fixed by the measurement of $I_V(\omega)$ and $I_H(\omega)$ in Fig.~\ref{fig:coinc}b. The $F$ parameter is set to $2$ to best reproduce the data within a range of values constrained by the material's anisotropy \cite{supp}. The measured and theoretical values demonstrate a good agreement. 
The Bell inequality for $S^{\Psi+}(\omega)$ is violated by up to $8.7$ standard deviations over a broad spectral range below 1250~cm$^{-1}$. 
The measured and theoretical values of $S^{\Psi-}(\omega)$ display no violation of the corresponding Bell inequality. This is a consequence of the sub-optimal spectral resolution of the measurement. In the ideal case, before the convolution with the Gaussian $G(\omega)$ (dashed black curve), the inequality is violated in a narrow spectral region around $1375\text{~cm}^{-1}$, where the biphoton state is $\ket{\psi(\omega)}\approx\ket{\Psi_-}$.
%Fig.~\ref{fig:interf}c reports measured and theoretical two-photon interference curves at $910\text{~cm}^{-1}$ for different sets of $(\theta_1,\theta_2)$. The correlation parameter for $\theta_1=\frac{\pi}{4}$ shows a small drop in visibility of $\simeq3.8\%$ compared to the theoretical curve from eq.~(\ref{eq:state}), attributed to imperfections of the experimental setup that reduce the purity of the generated quantum state.
%The dashed line is the sinusoidal fit $a \cos(2\theta_2-\frac{\pi}{2})$ with $a=0.85$.

\textit{Discussion.---} 
In our experiment, the coincidence detection rate of polarization-entangled photon pairs per unit bandwidth is of $\sim 0.25$~Hz/nm for a pump power of $48$~mW, with an overall detection efficiency not better than a few percent (including total internal reflection inside the diamond, fiber coupling losses, propagation losses and detector efficiency). The pair generation rate is therefore on the order of 10~Hz/nm for an interaction length of $\sim 3~\mu$m determined by the depth of focus. How much could the brightness of the source be improved? 
As detailed in the End Matter, based on the results from Ref.~\citenum{hausmann_diamond_2014}, we estimate that a centimeter-long diamond waveguide pumped under similar conditions would generate pairs at rate of 10 to 100~kHz/nm, which would further increase quadratically with pump power and be enhanced using photonic resonators. Our results therefore motivate further development of diamond photonic integrated circuits tailored to produce broadband polarization entangled photons around any center wavelength in the material's transparency window.   
% we model everything with a pure state
% we can increase our pump power and the counts will scale quadratically
%In \cite{fang_multidimensional_2016} frequency-resolved tomography is performed with stimulated emission tomography. There the polarization-entangled photons are also generated by FWM in an optical fiber with a Sagnac loop.

 %The main advantage of such systems over bulk materials, besides the potential for scalability, is that the phase matching conditions are relaxed along the spatial directions in which the structure is considerably smaller than the excitation wavelength. As a consequence, photon pairs are generated in a broad spectral and angular range, without restriction to specific temperature conditions \cite{sharapova_nonlinear_2023}. In addition, if the material structure holds one or more resonances overlapping with the correlated photons spectrum, the emission rate can be strongly enhanced. 

\subsection*{Data availability statement} 
The data that support the findings of this study will be deposited in a Zenodo repository with DOI...

\subsection*{Code availability statement} 
The code used in this study for data analysis and modeling will be deposited in a Zenodo repository with DOI...

\bibliography{biblio}

\subsection*{Acknowledgements}
We acknowledge Prof. Ado Jorio and his team for insightful discussions.
This work has received funding from the European Union's Horizon 2020 research and innovation program under Grant Agreement No. 820196 (ERC CoG QTONE), and from the Swiss National Science Foundation (project numbers 214993 and 198898).

\subsection*{Author contributions}
V.V. and F.C. performed the measurements, V.V. and S.P.A. implemented the fiber spectroscopy setup, V.V. analyzed the data and developed the model, V.V. and C.G. wrote the paper, C.G. led the project.

\subsection*{Competing interests}
The authors declare no competing interests.

\newpage
\section*{End Matter and Supplementary Material}

\paragraph{Arbitrary pump polarization. ---}
Let's consider a pump field with arbitrary polarization represented by the normalized amplitudes $a_V = \cos{\phi}$ and $a_H = \sin{\phi} e^{i \xi}$, where $\phi \in [0,\frac{\pi}{4}]$ and $\xi \in [0,\frac{\pi}{2}]$.
In addition to Eq.~\ref{chi_VVVV} and \ref{chi_HVVH}, we must consider 
\begin{equation}
    \chi_{VVHH}^{(3)} = \chi_{HVVH}^{(3)\text{E}} F' \left(1+\frac{1}{2F'}\frac{\Delta}{\Omega_\nu-\omega + i \Gamma_\nu} \right) \label{chi_VVHH}
\end{equation}
where $F' \equiv \chi_{VVHH}^{(3)\text{E}}/\chi_{HVVH}^{(3)\text{E}}$ is related to $F$ by the material's anisotropy \cite{supp} and the following identities hold
\begin{align}
    \chi_{VVVV}^{(3)} &= \chi_{HHHH}^{(3)} \\
    \chi_{HVVH}^{(3)} &= \chi_{VHHV}^{(3)}  \\
    \chi_{VVHH}^{(3)} &= \chi_{HHVV}^{(3)} =\chi_{HVHV}^{(3)} = \chi_{VHVH}^{(3)}. \label{chi_id}
\end{align}

Following Eqs.\ref{chi_VVVV}-\ref{chi_id}, we can write the non-normalized biphoton wavefunction generated by the nonlinear interaction for an arbitrary pump polarization
\begin{widetext}
\begin{align}
    \ket{\psi(\omega)} = \chi_{HVVH}^{(3)\text{E}} \Bigg[& \left(a_V^2F+a_H^2\left(1+\frac{\Delta}{\Omega_\nu-\omega + i \Gamma_\nu}\right)\right) \ket{VV} \nonumber\\
    &+\left(a_H^2F+a_V^2\left(1+\frac{\Delta}{\Omega_\nu-\omega + i \Gamma_\nu}\right)\right) \ket{HH} \label{eq:state_full}\\
    &+2a_Va_HF'\left(1+\frac{1}{2F'}\frac{\Delta}{\Omega_\nu-\omega + i \Gamma_\nu}\right) \left(\ket{HV}+ \ket{VH}\right)
    \Bigg].\nonumber   
\end{align}
When the Stokes and anti-Stokes polarizers are set at angles $\theta_1$ and $\theta_2$, respectively, from the vertical, the coincidence detection probability is computed as 
\begin{equation}
    P_{\theta_1,\theta_2}(\omega)=\frac{n_{\theta_1,\theta_2}(\omega)}{n_{\theta_1,\theta_2}(\omega)+n_{\theta_1+\pi/2,\theta_2+\pi/2}(\omega)+n_{\theta_1,\theta_2+\pi/2}(\omega)+n_{\theta_1+\pi/2,\theta_2}(\omega)}.\label{eq:P}
\end{equation} 
\end{widetext}
where 
\begin{equation}
    n_{\theta_1,\theta_2}(\omega)=P_{\theta_1,\theta_2}^\text{th}(\omega) \ast G(\omega) \label{eq:n}
\end{equation}
is the expected measured coincidence counts obtained by spectral convolution of the non-normalized theoretical joint detection probability $P_{\theta_1,\theta_2}^\text{th}(\omega)$ with the Gaussian modeling the setup response.
%$P_{\theta_1,\theta_2}^\text{th}(\omega)=|\braket{\theta_1,\theta_2 |\psi(\omega)}|^2$ (with $\ket{0}\equiv\ket{V}$) 
Note that $I_V(\omega)\propto P_{0,0}^\text{th}(\omega)$ and $I_H(\omega)\propto P_{\pi/2,\pi/2}^\text{th}(\omega)$ for $\phi=\xi=0$. From Eq.~\ref{eq:P}, we determine the correlation parameter
\begin{equation}
    E_{\theta_1,\theta_2}(\omega) = P_{\theta_1,\theta_2}(\omega) +P_{\theta_1+\pi/2,\theta_2+\pi/2}(\omega)-P_{\theta_1,\theta_2+\pi/2}(\omega)-P_{\theta_1+\pi/2,\theta_2}(\omega)\label{eq:E}
\end{equation}

\begin{figure}[h!]
\centering
\includegraphics[width=0.4\textwidth]{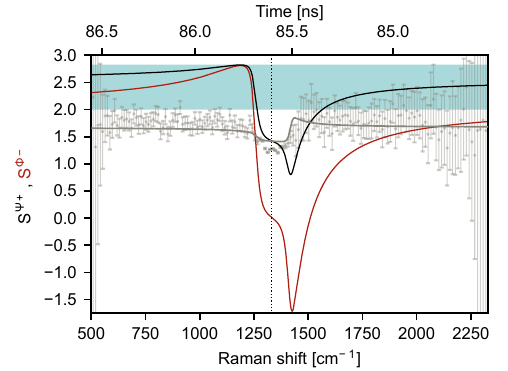}
\caption{{\bf Bell test for different pump polarizations.} Shaded gray circles represent the measured CHSH parameter $S^{\Psi+}(\omega)$ for linear pump polarization at $\phi = \frac{\pi}{4}$. The gray line is the corresponding theoretical curve. The black line represents the theoretical $S^{\Psi+}(\omega)$ for linear pump polarization at $\phi=0$ (vertical) as in Fig.~\ref{fig:interf}(b). The red line represents the theoretical CHSH parameter $S^{\Phi-}(\omega)$ for circular pump polarization, i.e. $\phi=\frac{\pi}{4}$ and $\xi=\frac{\pi}{2}$. The dotted black vertical line indicates the Raman resonance. The blue region spans between the classical upper bound of $2$ and the quantum upper bound of $2\sqrt{2}$. }
\label{fig:S_pol}
\end{figure}

In Fig.~\ref{fig:S_pol} we compare CHSH parameters computed for different pump polarizations to demonstrate that our measurement configuration ($\phi=\xi=0$) exhibits the maximum and broadest violation of Bell inequalities. Indeed, rotating the linear pump polarization from $\phi=0$ (vertical) to $\phi=\pi/4$ gradually degrades the CHSH parameter $S^{\Psi+}(\omega)$ in the off-resonance regions. To further validate our model, we experimentally implement a Bell test for the “worst" case $\phi=\pi/4$. In this configuration, the mixed terms of Eq.~\ref{eq:state_full} (proportional to $\ket{VH}$ and $\ket{HV}$) are non-zero. The measured $S^{\Psi+}(\omega)$ values shown in Fig.~\ref{fig:S_pol} manifest good agreement with the theoretical curve. 
We notice that a circular pump polarization ($\phi=\frac{\pi}{2}$, $\xi=\frac{\pi}{4}$) would also generate entangled photon pairs, albeit within a narrower spectral range than in the linear vertical case. Indeed, the computed CHSH parameter
\begin{equation}
    S^{\Phi-}(\omega)=-E_{0,\theta}(\omega)-E_{0,-\theta}(\omega)+ E_{2\theta,\theta}(\omega)- E_{2\theta,-\theta}(\omega)
\end{equation}
for circular polarization shows maximum violation of the corresponding Bell inequality ($S^{\Phi-}(\omega)\leq 2$) at low Raman shifts, where the Bell state $\ket{\Phi_-}=\frac{\ket{HV}-\ket{VH}}{\sqrt{2}}$ can be generated. 
%In Fig.~\ref{fig:S_pol}a we display the computed CHSH parameter $S^{\Psi+}$ for different linear and elliptical pump polarizations, demonstrating that our measurement configuration ($\phi=\xi=0$) exhibits the maximum violation of the Bell inequality. To further validate our model, we experimentally implement a Bell test with linear pump polarization at $\phi=\frac{\pi}{4}$. In this configuration, the mixed terms of Eq.~\ref{eq:state_full} (proportional to $\ket{VH}$ and $\ket{HV}$) are non-zero. The measured $S$ values shown in Fig.~\ref{fig:S_pol}b manifest good agreement with the theoretical curve. 
%The opaque black and red lines represent the theoretically computed CHSH parameter $S$ for a linearly polarized pump at $\phi=0$ (vertical, cf. Fig.~\ref{fig:interf}(b)) and $\phi=\frac{\pi}{4}$ respectively. Black lines with increasing transparency indicate increasing polarization angles from $\phi=0$ to $\simeq \frac{\pi}{4}$. The red lines are all computed at $\phi=\frac{\pi}{4}$ with transparency increasing with the phase shift angle from $\xi = 0$ (linear polarization) to $\xi =\frac{\pi}{2}$ (circular polarization).

\paragraph{Improving the photon pair generation rate.---}
We estimate here the photon pair generation rates that could be achieved using diamond waveguides. 
First, the wave-vector mismatch, defined as $\Delta k_L = 2k_p - k_S - k_A$, where $k_{p/S/A}$ are the pump, Stokes and anti-Stokes wave-vectors, fixes an intrinsic coherence length of $L_\text{coh}=\frac{2\pi}{\Delta k_L} \gtrsim 700$~$\mu$m for diamond's dispersion around 780~nm. 
Second, Ref.~\citenum{hausmann_diamond_2014} shows that the coherence length can be increased at will by suitably engineered diamond waveguides. Indeed, FWM in waveguides requires satisfying the phase-matching condition $\Delta k = 2\gamma P_p -\Delta k_L\approx 0$, where $P_p$ is the pump power, $\gamma=\omega_p n_2/cA_\text{eff}$ is the effective nonlinearity of the waveguide, with $n_2$ the nonlinear refractive index ($\approx 1.23 \times 10^{-19}$~m$^2$/W) and $A_\text{eff}$ the effective mode area. The waveguide can be engineered to have $\Delta k_L > 0$, i.e. an anomalous (positive) group velocity dispersion $D$ around the pump wavelength. 
As a consequence, a pump power of $P_p = \Delta k_L/2\gamma $ gives in principle an infinite coherence length over some bandwidth. Simulations show that $D\approx 500$~ps/(nm Km) at an excitation wavelength of $800$~nm for $A_\text{eff}\approx 400 \times 500$~nm$^2$ \cite{hausmann_diamond_2014}. Since $\Delta k_L = D {\omega_p^2}2\pi c(\omega_p-\omega_s)^2/\omega_p^2$, we calculate an optimal pump power of $P_p< 1$~kW that has the same order of magnitude as the peak power we employ in our experiment (which is $\sim 600$~W). For a centimeter-long waveguide pumped in these conditions, the estimated photon pair generation rate is on the order of 10 to 100~kHz/nm. Of course, in such a realization, the TE/TM birefringence of the waveguide should also be minimized.

\paragraph{Fiber spectroscopy calibration.---} 
In order to extract spectral information from the arrival time of the photons on the photodetectors, it is necessary to calibrate the fiber spectroscopy technique. The single-mode fiber collecting the anti-Stokes signal is an S630-HP with a single-mode bandwidth of $[630-860]$~nm and length $L_A = 25$~m. The Stokes signal travels through two 780-HP fibers with a single-mode bandwidth of $[780-970]$~nm for a total length $L_S = 125$~m. The dispersion parameters $D_{S/A}(\omega)$ of each fiber are taken from the datasheets. 
\begin{figure}[h!]
\centering
\includegraphics[width=0.5\textwidth]{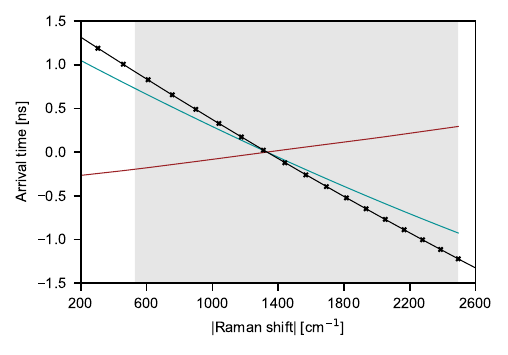}
\caption{{\bf Time-shift calibration.} The red and blue lines indicate the arrival times of the Stokes photons $T_S$ and of the anti-Stokes photons $T_A$ respectively as a function of the absolute Raman shift. Black crosses represent the calculated coincidence times $T_C$; the black line is the corresponding fit. The gray area denotes the spectral region in which correlated Stokes and anti-Stokes photons can be detected, and it's limited by the fibers' bandwidths and the spectral filters.}
\label{fig:SM_cal}
\end{figure}
The arrival times of the Stokes and anti-Stokes photons are then calculated as
\begin{equation}
    T_{S/A}(\omega)=-2\pi c L_{S/A} \int  \frac{D_{S/A}(\omega)}{\omega^2} d\omega.
\end{equation} 
The time-to-digital converter is configured to receive the anti-Stokes signal as “start" input and the Stokes signal as “stop" input. Therefore, the coincidence time delay is defined by 
\begin{equation}
    T_C(\omega)=T_S(\omega)-T_A(\omega-2\omega_L).
\end{equation}
Fig.~\ref{fig:SM_cal} shows $T_S$, $T_A$ and $T_C$ as a function of the absolute Raman shift with the time axis arbitrarily centered at the Raman resonance ($1332\text{~cm}^{-1}$). For simplicity, the calculated values of $T_C$ are fitted with the function $T_\text{fit}(\omega)= a+b\cdot \omega^2$, since the arrival time is inversely proportional to the group velocity $v_g=c/(n+\omega\frac{dn}{d\omega})$ and the refractive index can be expanded as $n\simeq A+ B\cdot\omega^2$. 
%where $a=-6.35$~ns and $b=1.36\times 10^{-6}$~ns/(rad THz)
%can be rewritten as $T=\frac{L}{v_g}$
%the refractive index can be expanded as $n\simeq A+ B\omega^2$ and $ T= \frac{L}{v_g(\omega)}=\frac{L}{c}(n+\omega\frac{dn}{d\omega})$, $v_g$ being the group velocity.
The measurements reported here and in the main text are calibrated by shifting the curves $T_S$, $T_A$ and $T_C$ along the time axis to center the Raman peak at $1332\text{~cm}^{-1}$.

\paragraph{Diamond orientation.---} 
The $\chi^{(3)}$ tensor is a sum of electronic and Raman contributions with different weights depending on the field polarization relative to the crystal axes. 
When the pump polarization is aligned with the [001] axis of diamond, the Stokes and anti-Stokes photons generated through the exchange of a real phonon exhibit polarizations orthogonal to the pump. Consequently, the frequency difference $\Delta$ between the Raman peak and the point of maximum destructive interference reaches its maximum value \cite{levenson_dispersion_1974} (cf. Fig.~\ref{fig:coinc}).
To realize this configuration, the diamond crystal is cut perpendicular to the [100] crystal axis and is mounted on a rotation stage. Since the pump polarization is fixed to vertical in the lab frame, the crystal is rotated to maximize the horizontally polarized Stokes power in order to align the [001] axis along the pump polarization (cf. Fig.~\ref{fig:setup}). 

%The LP and SP filters are chosen such that to obtain the right cut for the pump spectrum we can tilt them with the same angle in opposite directions. In this way, any tiny horizontal component to the initial polarization, which can be present because 
%diamond sample—grown along the [100] direction by a high-pressure, high-temperature method, about 300 mm thick and polished on both faces along the (100) crystallographic plane

\begin{figure}[h!]
\centering
\includegraphics[width=0.5\textwidth]{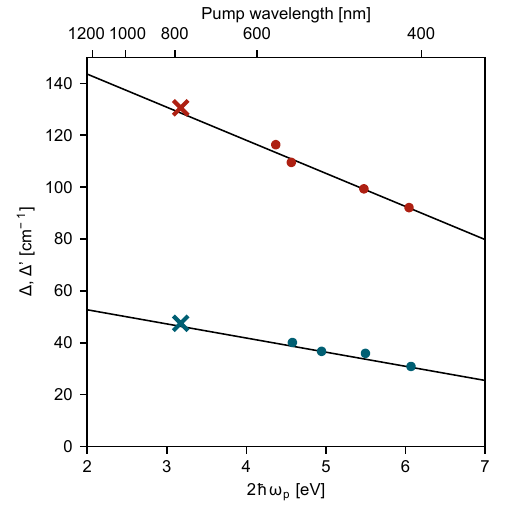}
\caption{{\bf Dependence of $\Delta$ and $\Delta'$ on the excitation wavelength.} The red and blue circles represent, respectively, the measurements of $\Delta$ and $\Delta'$ reported in Ref.\cite{levenson_dispersion_1974}. The black lines are the corresponding linear regressions. The red cross indicates the value $\Delta=130.59\text{~cm}^{-1}$ that we measure at $781$~nm. The blue cross indicates the value $\Delta'=47.49\text{~cm}^{-1}$, calculated by inserting the measured $\Delta$ and the parameters $F=2$ and $F'=1.25$ in Eq.~\ref{eq:Delta3}, showing that the same linear trend vs. wavelength as in Ref.\cite{levenson_dispersion_1974} is satisfied.}
\label{fig:Delta}
\end{figure}
\textit{Diamond anisotropy.---} 
In a perfectly isotropic material the relationship between the electronic components of the susceptibility tensor should follow $\chi_{VVVV}^{(3)\text{E}}\simeq3\chi_{HVVH}^{(3)\text{E}}\simeq3\chi_{VVHH}^{(3)\text{E}}$. With our definition of $F$ and $F'$ in the main text this relationship translates into $F\simeq3$, $F'\simeq1$; we can thus define an anisotropy parameter
\begin{align}
    \sigma_a &\equiv\frac{2\chi_{VVHH}^{(3)\text{E}}+\chi_{HVVH}^{(3)\text{E}}-\chi_{VVVV}^{(3)\text{E}}}{\chi_{VVVV}^{(3)\text{E}}} \nonumber \\
    &=\frac{2F'+1-F}{F}.
\end{align}
In diamond, it is expected that $\sigma_a$ is small but non-zero because the electronic orbitals responsible for the third-order nonlinearity are not completely spherical.
Moreover, the values of $F$, $F'$ and $\sigma_a$ may change between different diamond samples depending on the presence of strain and defects, and they may depend on the excitation wavelength. 
In principle, the values of $F$ and $F'$ can be extracted by coincidence measurements far from the Raman resonance, but the polarization response of the setup and the limited frequency range of our detection scheme affect the precision of this approach.
Alternatively, the variation of the anisotropy parameters with wavelength can be deduced from the measurement of $\Delta$ and
\begin{equation}
    \Delta' = \frac{2 \Delta}{F+1+2F'} = \frac{2 \Delta}{F(\sigma_a+2)} \label{eq:Delta3}
\end{equation}
reported by Ref. \cite{levenson_dispersion_1974} at various excitation wavelengths between $400$ and $600$~nm, and reproduced in Fig.~\ref{fig:Delta} with full circles. %from which we can extract the following parameters at $781$~nm: $\Delta\simeq130$~cm$^{-1}$, $\Delta'=47.49$, $F\simeq 2.6$, and $F'\simeq 1$ for a reported $\sigma_a=0.17$. 
The parameter $\Delta'$ represents the frequency difference between the Raman peak and the point of maximum destructive interference when the pump polarization is diagonal at $\phi=\frac{\pi}{4}$ and the polarizers in detection are parallel to the pump.

Fig.~\ref{fig:Delta} shows that the value of $\Delta=130.59\text{~cm}^{-1}$ we measure at $781$~nm (see main text) is in agreement with the measurements of Ref. \cite{levenson_dispersion_1974}. Therefore, we can choose the parameters $F$ and $\sigma_a$ to obtain a value of $\Delta'$ compatible with the same measurements. We select $F=2$, since larger values lead to a drop in the $S$ parameter at low Raman shifts, where our measurements demonstrate $S\simeq 2\sqrt{2}$. From Eq.~\ref{eq:Delta3}, with $\Delta'=47.49\text{~cm}^{-1}$ compatible with Fig.~\ref{fig:Delta} \cite{levenson_dispersion_1974}, we calculate an anisotropy factor of $\sigma_a= 0.75$, and hence $F' = 1.25$. These parameters are used for all the theoretical curves. 
% polarization response of the setup

\begin{figure*}[h!]
\centering
\includegraphics[width=1\linewidth]{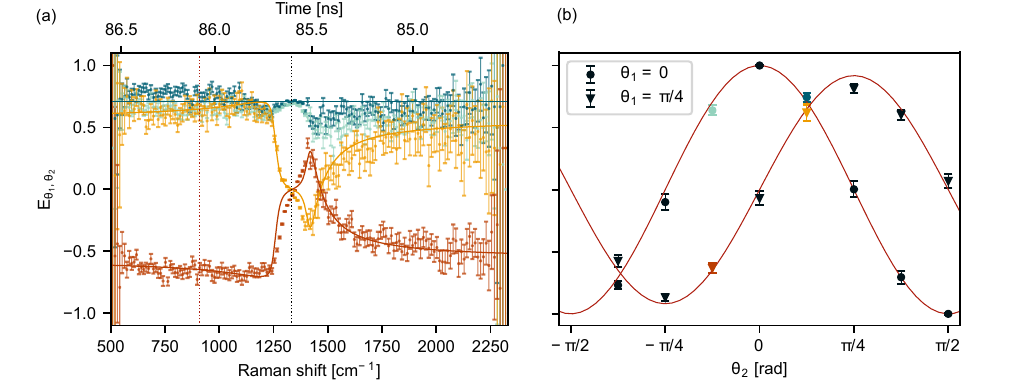}
\renewcommand{\baselinestretch}{1.1}
\caption{{\bf Bell test measurements and two-photon interference.} (a) Measured values of $E_{\theta_1,\theta_2}(\omega)$ for a vertically polarized pump are shown as dark cyan circles for the angles $(\theta_1,\theta_2) = (0,\frac{\pi}{8})$, light cyan for $(0,-\frac{\pi}{8})$, yellow for $(\frac{\pi}{4},\frac{\pi}{8})$ and orange for $(\frac{\pi}{4},-\frac{\pi}{8})$. The dotted black vertical line indicates the Raman resonance. The error bars are calculated by propagating the standard deviation of the coincidence counts in the region $[630, 890]\text{~cm}^{-1}$. (b) Measured $E_{\theta_1,\theta_2}$ at $910\text{~cm}^{-1}$ (dotted red line in (a)) for $\theta_1=0$ (circles) and $\theta_1=\frac{\pi}{4}$ (triangles) and different values of $\theta_2$. Colored points for the same angles as (a). In all panels, the solid lines are theoretical curves computed from Equations \ref{eq:n}-\ref{eq:S}. }
\label{fig:interf_suppl}
\end{figure*}

\paragraph{Bell test measurements.---}
Fig.~\ref{fig:interf_suppl} clarifies the relationship between the Bell test and the two-photon interference measurements presented in the main text. The Bell test requires the measurements of the correlation parameter $E_{\theta_1,\theta_2}(\omega)$ for the $4$ couples of detection angles $(\theta_1,\theta_2)=(0,\theta)$, $(0,-\theta)$, $(2\theta,\theta)$, and $(2\theta,-\theta)$ where $\theta=\pi/8$ (cf. panel (a)). For each couple of angles, we perform $4$ coincidence measurements $n_{\theta_1',\theta_2'}(\omega)$ of $1$ hour with the Stokes and anti-Stokes polarizers set to $(\theta_1',\theta_2')=(\theta_1,\theta_2)$, $(\theta_1+\pi/2,\theta_2+\pi/2)$, $(\theta_1,\theta_2+\pi/2)$, and $(\theta_1,\theta_2+\pi/2)$. 
Then, the correlation parameter is computed from Eq.~\ref{eq:P} and \ref{eq:E}. Thereby, the CHSH parameters $S^{\Psi\pm}(\omega)$ computed from Eq.~\ref{eq:S} and shown in Fig.~\ref{fig:interf} are the result of $16$ coincidence measurements. The error bars are calculated by propagating the standard deviation of the coincidences in the region $[630, 890]\text{~cm}^{-1}$, where the counts are almost constant. To demonstrate two-photon interference, we perform $12$ additional measurements of $E_{\theta_1,\theta_2}(\omega)$ for different polarization angles $(\theta_1,\theta_2)$ (cf. panel (b)). 
We notice that $16$ data points generally require $64$ coincidence measurements, although our choices of $(\theta_1,\theta_2)$ reduce this number to $44$.

\end{document}